\documentclass[journal=jacsat,manuscript=article]{achemso}
\usepackage{epstopdf}
\usepackage[normalem]{ulem}
\usepackage{xcolor} % Required for color customization
\usepackage[colorlinks]{hyperref}

\hypersetup{
    colorlinks=true,
    linkcolor=blue, % Set color for links (change as needed)
    urlcolor=blue, % Set color for URLs (change as needed)
    citecolor=blue, % Set color for citations (change as needed)
}
\AppendGraphicsExtensions{.tif}

\usepackage[version=3]{mhchem} % Formula subscripts using \ce{}https://www.overleaf.com/project/65415c7a44af46c3f1ea9450

\author{Muhammad Hassan Shaikh}
\affiliation{Department of Physics, University of Delaware, Newark, Delaware, 19716, USA}
\author{Matthew Whalen}
\affiliation{Quantum Science and Engineering Program, University of Delaware, Newark, Delaware, 19716, USA}
\author{Dai Q. Ho}
\affiliation{Department of Material Science and Engineering, University of Delaware, Newark, Delaware, 19716, USA}
\alsoaffiliation{Faculty of Natural Sciences, Quy Nhon University, Quy Nhon 55113, Vietnam}
\author{Aqiq Ishraq}
\affiliation{Department of Material Science and Engineering, University of Delaware, Newark, Delaware, 19716, USA}
\author{Collin Maurtua}
\affiliation{Department of Material Science and Engineering, University of Delaware, Newark, Delaware, 19716, USA}
\author{Kenji Watanabe}
\affiliation{National Institute for Materials Science (NIMS) in Tsukuba, 305-0047, Japan}
\author{Takashi Taniguchi}
\affiliation{National Institute for Materials Science (NIMS) in Tsukuba, 305-0047, Japan}
\author{Yafei Ren}
\affiliation{Department of Physics, University of Delaware, Newark, Delaware, 19716, USA}
\author{Anderson Janotti}
\affiliation{Department of Material Science and Engineering, University of Delaware, Newark, Delaware, 19716, USA}
\author{John Xiao}
\affiliation{Department of Physics, University of Delaware, Newark, Delaware, 19716, USA}
\alsoaffiliation{Quantum Science and Engineering Program, University of Delaware, Newark, Delaware, 19716, USA}
\author{Chitraleema Chakraborty}
\affiliation{Department of Physics, University of Delaware, Newark, Delaware, 19716, USA}
\alsoaffiliation{Quantum Science and Engineering Program, University of Delaware, Newark, Delaware, 19716, USA}
\alsoaffiliation{Department of Material Science and Engineering, University of Delaware, Newark, Delaware, 19716, USA}
\email{cchakrab@udel.edu}
\phone{+1(302) 831-3251}

\title
  {\textbf{Magnetic proximity coupling to defects in a two-dimensional semiconductor}}


\begin{document}

% \begin{tocentry}

% Some journals require a graphical entry for the Table of Contents.
% This should be laid out ``print ready'' so that the sizing of the

% Inside the \texttt{tocentry} environment, the font used is Helvetica
% 8\,pt, as required by \emph{Journal of the American Chemical

% The surrounding frame is 9\,cm by 3.5\,cm, which is the maximum
% permitted for  \emph{Journal of the American Chemical Society}
% graphical table of content entries. The box will not resize if the
% content is too big: instead it will overflow the edge of the box.

% This box and the associated title will always be printed on a
% separate page at the end of the document.

% \end{tocentry}

\begin{abstract}
  The ultrathin structure and efficient spin dynamics of two-dimensional (2D) antiferromagnetic (AFM) materials hold unprecedented opportunities for ultrafast memory devices, artificial intelligence circuits, and novel computing technology. For example, chromium thiophosphate (CrPS$_4$) is one of the most promising 2D A-type AFM materials due to its robust stability in diverse environmental conditions and net out-of-plane magnetic moment in each layer, attributed to anisotropy in crystal axes (a and b). However, their net zero magnetic moment poses a challenge for detecting the N\'eel state that is used to encode information. In this study, we demonstrate the detection of the N\'eel vector by detecting the magnetic order of the surface layer by employing defects in tungsten diselenide (WSe$_2$). These defects are ideal candidates for optically active transducers to probe the magnetic order due to their narrow linewidth and high susceptibility to magnetic fields. We observed spin-polarized charge transfer in the heterostructure of bulk CrPS$_4$ and single-layer WSe$_2$ indicating type-II band alignment as supported by density functional theory (DFT) calculations. In the A-type AFM regime, the intensity of both right-handed and left-handed circularly polarized light emanating from the sample remains constant as a function of the applied magnetic field, indicating a constant polarized transition behavior. Our results showcase a new approach to optically characterizing the magnetic states of 2D bulk AFM material, highlighting avenues for future research and technological applications. 
 
\end{abstract}


Replacing ferromagnetic (FM) materials in commercial magnetic random-access memories (MRAMs) \cite{1,2} and other applications such as artificial intelligence, neuromorphic computing, and in-memory computing\cite{2,3} with AFM material holds the potential to improve the operation speed (from GHz to THz), the immunity to electromagnetic interference, and the memory density \cite{4,5,6,7}. 2D AFM materials further push the limit to atomic layer thickness. The challenges are also daunting to implement AFM material-based devices since AFM materials nearly lack physical observables (resistance, voltage, etc.) that are related to
their magnetic order parameter, known as the Néel vector \cite{2}. Bringing the evident advantages of AFM materials into play by being able to detect the Néel vector would allow using their unique functionalities in future generations of advanced technologies.

Recently, binary transition metal chalcogenides (TMCs) have garnered significant attention due to their layered-dependent magnetic properties. Due to the van der Waals forces between each layer, the material can be exfoliated down to a single layer that possesses a magnetic ground state which enables the switching between FM and AFM material behavior contingent upon the number of layers\cite{mcguire2020cleavable,basak2023theoretical}.  Chromium triiodide (CrI$_3$) is one of the binary TMCs that possess A-type AFM, with FM ordering within the same layer and AFM 
between adjacent layers. Consequently, thin flakes of a few layers can display net or zero magnetic moment depending on whether there is an odd or even number of layers. The surface layer magnetic moment was successfully detected in the A-type AFM phase by utilizing another 2D optically active material WSe$_2$ atop CrI$_3$\cite{surfacelayercri3,wei2019recent}. WSe$_2$ possesses valley-selective transition and in the presence of a net out-of-plane magnetic field, the valley degeneracy can be lifted. Due to the high susceptibility of WSe$_2$ to the magnetic field, the surface layer magnetic moment of a magnetic material can be detected by measuring the energy splitting and observing differences in the intensity of valley transitions in WSe$_2$. However, CrI$_3$ is not stable in an open environment which creates a problem in using this material in practical applications\cite{wei2019recent}.

Recently studied ternary TMCs CrPS$_4$ emerge as particularly noteworthy due to their stability in bulk form and high resonance capability, making them a viable candidate for advanced spintronic and memory device applications\cite{atypeafmairstable,spinflipandafm,yang2021layer}. Notably, CrPS$_4$ exhibits an anisotropic crystal structure endowing each layer with net magnetic moments that are antiferromagnetically coupled, resulting in the same A-type AFM behavior as CrI$_3$\cite{atypeafmairstable,magneticgroundstate,spinflipandafm}. This anisotropy extends to light absorption, rendering CrPS$_4$ suitable for polarization-based sensing and memory devices\cite{anisotropycrystal,polarizedraman}. Its N\'eel temperature is 37 K.

However, its A-type AFM state in a bulk form with many layers that are antiferromagnetically coupled makes it difficult to sense the small change in the surface layer magnetization. To overcome this limitation, we propose integrating CrPS$_4$ with defect-based bound excitons in TMDCs as a means to rectify surface layer magnetization, owing to their narrow linewidth relative to excitonic peaks, high excitonic \textit{g} factor, and efficient electrical tunability \cite{chakraborty2017quantum,
chakraborty20183d,
palacios2018atomically,
chakraborty2015voltage, hybridization, nanobubble,chakraborty2019advances,moon2020dynamic,chakraborty2016localized,moon2020strain}. Additionally, we present a method for detecting A-type AFM and canted antiferromagnetic (C-AFM) states using photoluminescence (PL) light polarization of bulk CrPS$_4$, providing insights into its magnetic behavior. We also observed constant polarized transition behavior from CrPS$_4$ in the A-type AFM regime. This constant polarized transition holds promise for potential applications in spintronics and ultrafast AFM-based memory devices. Our method not only enables the detection of surface layer magnetic moments but also provides valuable information about the various phases of antiferromagnetism present within bulk CrPS$_4$. This comprehensive understanding of the material's magnetic behavior holds immense promise for advancing a wide range of technological applications, from data storage to tunable optical devices, and beyond\cite{wei2019recent}.

\section{Results and discussion}

%first figure
\begin{figure}[!ht]
    \includegraphics[width=0.95\textwidth]{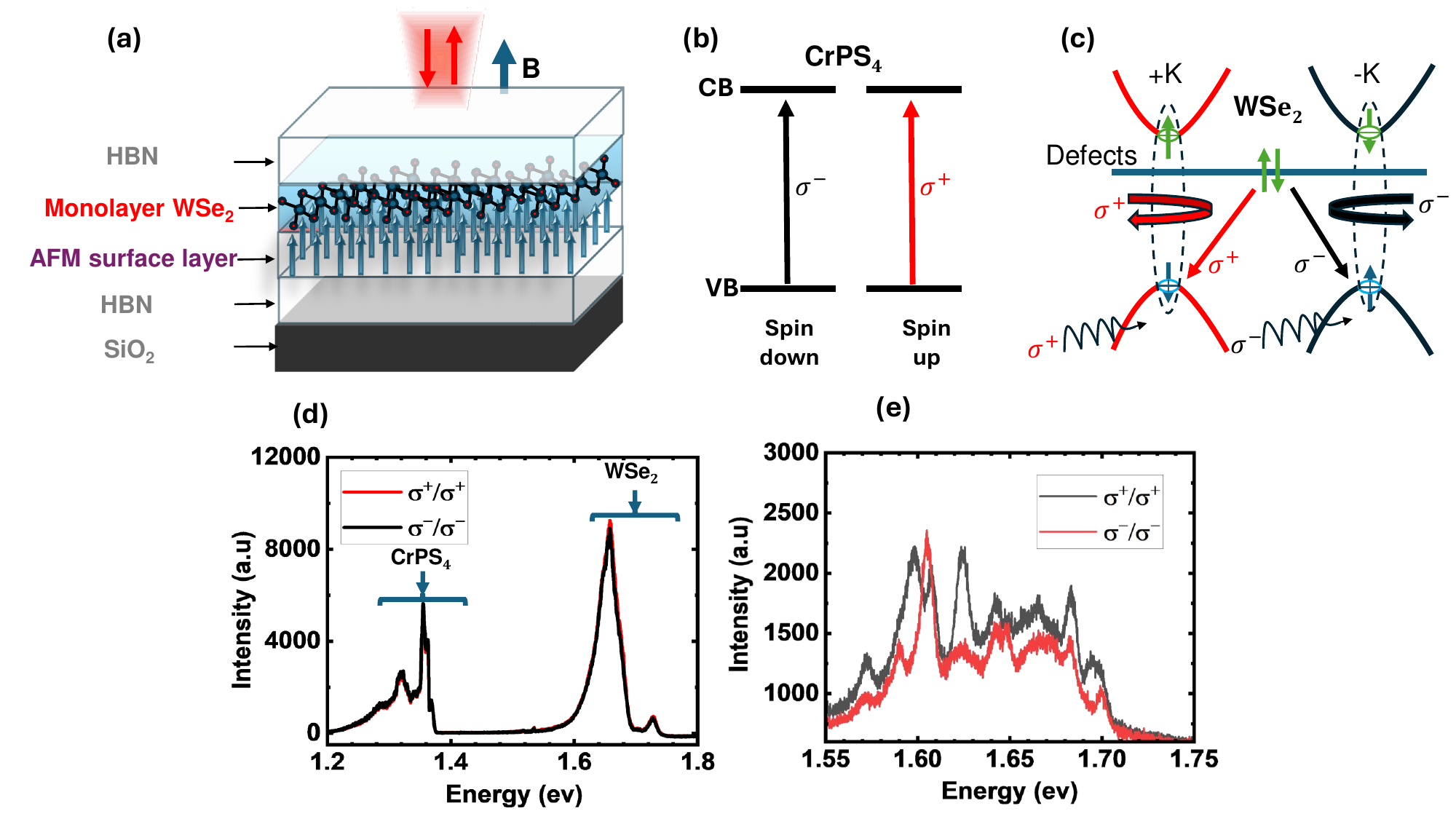}
    \caption{Low-temperature polarization-resolved photoluminescence (PL) of the CrPS$_4$ and WSe$_2$ heterostructure: In (a), a schematic illustrates the magnetic proximity interaction in a 2D system. Optical selection rules for CrPS$_4$ and WSe$_2$  are shown in (b) and (c), respectively. In (d), the first sample co-polarization resolved PL of the heterostructure reveals multiple peaks between 1.2-1.4 eV attributed to CrPS$_4$ due to vibrational progression. The peak centered around 1.65 eV corresponds to intrinsic defect states PL in WSe$_2$, and the 1.725 eV peak indicates the excitonic transition in WSe$_2$ at 1.8 K, (e) shows the second sample co-polarization resolved PL of defect-based bound excitons of WSe$_2$ within the CrPS$_4$ and WSe$_2$ heterostructure at 1.8 K .}
\end{figure}

In this study, we investigated the magnetic proximity interaction (MPI) between bulk CrPS$_4$ and single-layer WSe$_2$. Fig. 1(a) shows a schematic of the magnetic heterostructure utilized for studying this interaction, where the magnetic material is closely interfaced with the optically active 2D semiconductor. The properties of light emitted by the 2D semiconductor depend on the orientation of the surface layer magnetic moments within the magnetic material. This magnetic heterostructure could be utilized for potential ultrafast optical transduction. Both bulk CrPS$_4$ (the magnetic material) and single-layer WSe$_2$ (optically active semiconductor) were mechanically exfoliated using the scotch tape method, and the heterostructure was assembled by stacking both flakes together employing a poly-carbonate (PC) assisted transfer technique \cite{PCtransfer}, with detailed procedures outlined in the methods section.

We conducted co-polarization resolved photoluminescence (PL) measurements on the sample, where the sample was excited with circularly polarized light ($\sigma^+$ or $\sigma^-$) and the emitted circularly polarized light ($\sigma^+$ or $\sigma^-$) was detected from the heterostructure. Room temperature co-polarization resolved PL spectra shown in Supplementary Figure (S) 1, wherein the peak centered around ~1.35 eV corresponds to CrPS$_4$, while the peak centered around ~1.65 eV corresponds to excitonic transitions in WSe$_2$. Fig. 1(b) and Fig. 1(c) illustrate the optical selection rules for optical transitions in CrPS$_4$ and WSe$_2$. 

At room temperature, we observed no evidence of MPI due to the time-reversal symmetry as CrPS$_4$ remains in a paramagnetic phase. The valley degeneracy is only lifted in the presence of a magnetic field. 

 We cooled the sample down in a cryostat to 1.8 K, which is below the N\'eel temperature of CrPS$_4$ (37 K),\cite{spinflipandafm} and measured the co-polarization resolved PL of two different heterostructures(1(d) and 1(e)). Fig. 1(d) displays the PL of the heterostructure where the PL contributions from CrPS$_4$ and WSe$_2$ are indicated. As previously observed, the presence of multiple peaks in the CrPS$_4$ energy range can be attributed to a vibrational progression\cite{ISCincrps4}. The peak centered around ~1.65 eV corresponds to intrinsic defect states in WSe$_2$ which are defect states due to the impurities present in the material. The peak at ~1.725 eV corresponds to excitonic transitions in WSe$_2$. Due to the broad linewidth of the defect and excitonic transitions in Figure 1(d), it is challenging to resolve the valley splitting.

Additionally, we conducted the same experiment on a heterostructure consisting of CrPS$_4$ with defect-based bound excitons in WSe$_2$ (Fig. 1(e)). The defect-based bound excitons can arise from strain-induced wrinkles and nanobubbles formed during heterostructure preparation\cite{chakraborty2015voltage, hybridization, nanobubble,chakraborty2019advances,moon2020dynamic,moon2020strain}. The intrinsic defect states of WSe$_2$ hybridize with the optically dark excitonic states in the conduction band of WSe$_2$ due to strain, which creates localized potential within the energy band gap that funnels the excitons\cite{chakraborty2015voltage, hybridization, nanobubble,chakraborty2019advances,moon2020dynamic,moon2020strain}. This leads to the formation of multiple optically active defect-based bound excitonic states with very narrow linewidth compared to the intrinsic defect states. In Figure 1(e), the finite polarization contrast and energy splitting can be observed in the PL spectrum of defect-based bound excitons in WSe$_2$ even without the application of an external magnetic field.

The presence of zero-field splitting and finite polarization contrast in the PL spectrum indicates the signature of MPI between the surface layer of bulk CrPS$_4$ and single-layer WSe$_2$. While the net magnetic moment is zero in bulk CrPS$_4$. MPI being a short-range interaction, influences WSe$_2$ to experience a net out-of-plane magnetic moment from the surface layer of CrPS$_4$. This leads to a finite polarization contrast at low temperatures without the application of an external magnetic field. Such polarization contrast was previously observed only via the application of a Zeeman energy via an external magnetic field which overcomes the anisotropic Coulomb exchange interaction in localized excitons\cite{chakraborty2015voltage, hybridization, nanobubble,chakraborty2019advances}. We observe a similar polarization contrast that points to the restoration of the valley polarization inherited from the 2D excitons in optically active defects in 2D semiconductors\cite{chakraborty2015voltage, hybridization, nanobubble,chakraborty2019advances}

 %second figure
 \begin{figure}[!ht]

    \includegraphics[width=0.98\textwidth]{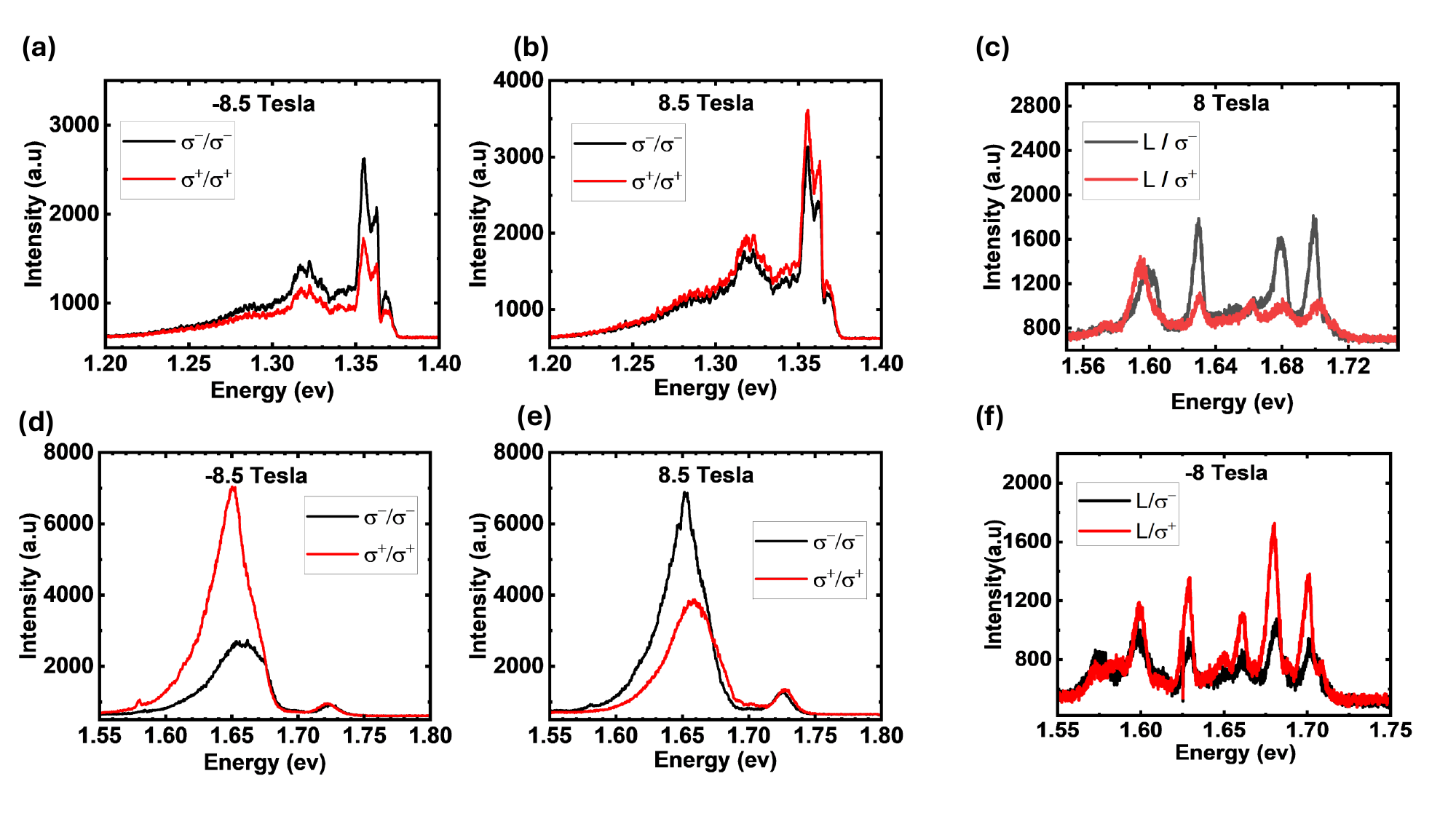}
    \caption{Magnetic field-dependent PL of the heterostructure in Faraday's geometry: (a) and (b) shows the PL contribution from CrPS$_4$ at high magnetic fields of $\pm$ 8.5 Tesla, while (d) and (e) show the PL contribution of intrinsic defects and excitons from WSe$_2$ under similar magnetic conditions. Additionally, (c) and (f) present the PL contribution of defect-based bound excitons from WSe$_2$ excited by linear polarized (L) light at high magnetic fields of $\pm$ 8 Tesla. }
\end{figure}

In Fig. 2, we conducted co-polarization resolved PL measurements at a high magnetic field in a Faraday geometry to further study the MPI effect in the heterostructure. Fig. 2(a) and 2(b) display the PL contributions of CrPS$_4$ at $\pm$ 8.5 T. We observed that when the applied field aligns with the crystal +c-axis, the PL contribution from CrPS$_4$ was dominated by $\sigma^+$ (Fig. 2(b)), as most of the magnetic moment of CrPS$_4$ aligns with the direction of the magnetic field. Conversely, when we switched the magnetic field direction and aligned it with the crystal -c-axis, the PL contribution from CrPS$_4$ was dominated by $\sigma^-$ (Fig. 2(a)), as most of the magnetic moment of CrPS$_4$ attempt to align with the -c-axis direction of the crystal.

In contrast, the PL contributions from intrinsic defect states and excitons in Fig. 2(d) and 2(e), and the PL contribution from the defect-based bound excitons in Fig. 2(c) and 2(f), exhibit opposite behavior compared to CrPS$_4$. When the B field aligns along the +c-axis (-c-axis), the PL contribution is dominated by $\sigma^-$ ($\sigma^+$)  from both intrinsic defect and defect-based bound excitons. We observed negligible change in the polarization contrast of excitonic transition in WSe$_2$ due to an increase in the defect-mediated valley scattering effect\cite{wagner2021trap}.
To further analyze the opposite behavior of the PL component from intrinsic defect states and defect-based bound excitons of WSe$_2$, we measured the degree of circular polarization ($\rho$) at different applied magnetic fields along Faraday's geometry, defined as:
\begin{equation}
\rho = \frac{I(\sigma^+/\sigma^+)-I(\sigma^-/\sigma^-)}{I(\sigma^+/\sigma^+)+I(\sigma^-/\sigma^-)}
\end{equation}

We conducted $\rho$ measurements at $\pm$ 8.5 Tesla (S2). S2(a) illustrates the PL range of CrPS$_4$, while S2(b) depicts the PL range of WSe$_2$. In CrPS$_4$, all peaks exhibit positive (negative) $\rho$ values at +8.5 (-8.5) Tesla due to anisotropy in the co-polarization component of PL. However, in the PL range of WSe$_2$, the peaks around 1.65 eV and 1.725 eV corresponding to intrinsic defects and excitonic transitions exhibit an opposite behavior compared to CrPS$_4$ at $\pm$ 8.5 Tesla. Nonetheless, certain regions (\textit{e.g.}, peaks at 1.68 eV and 1.73 eV) demonstrate similar behavior to CrPS$_4$, which is attributed to Zeeman splitting in WSe$_2$, as depicted in Fig. 2(d) and 2(e).

Fig. 3(a) displays $\rho$ as a function of the magnetic field of one of the peaks of CrPS$_4$ at 1.355 eV, while Fig. 3(b) shows the $\rho$ as a function of the magnetic field of the intrinsic defect peak of WSe$_2$ centered at 1.65 eV. Notably, the behavior of $\rho$ for CrPS$_4$ and WSe$_2$ exhibits an opposite trend in a high magnetic field.

We also observed that when CrPS$_4$ is in the A-type AFM state within the magnetic field range of $\pm$ 0.8 Tesla,\cite{spinflipandafm} the $\rho$ from CrPS$_4$ is non-zero but remains almost constant (Fig. 3(d)), whereas the $\rho$ from the defect peak of WSe$_2$ increases in a positive (negative) direction when the field is aligned along the negative (positive) c-axis of the crystal.

This behavior can be understood through the Type-II band alignment model between CrPS$_4$ and WSe$_2$, as shown in Fig. 3(f). According to this model, the specific spin orientation of electrons in the valleys of WSe$_2$ can transition to the spin selective state in the conduction band of CrPS$_4$. In the positive magnetic field scenario when most electrons of the surface layer of CrPS$_4$ align with the positive field, electrons at the +K valley of WSe$_2$ are transferred to the CrPS$_4$ conduction band path followed by an optical transition. This leads to a decrease in the $\sigma^+$ transition compared to $\sigma^-$ in WSe$_2$ defect states resulting in negative $\rho$ for WSe$_2$ defect states in a positive magnetic field. Conversely, when we switch the magnetic field direction, the behavior also switches accordingly.

In this manner, we detect the spin-polarized charge transfer effect between CrPS$_4$ and WSe$_2$ heterostructure. We also observe a similar effect on $\rho$ of defect-based bound excitons in WSe$_2$ as shown in the color plot in Fig. 3(c). This plot illustrates the negative (positive) values of $\rho$ for defect-based bound excitons in positive (negative) magnetic fields, indicating the same spin-polarized charge transfer effect observed in intrinsic defect states of WSe$_2$.

This effect was also observed in the heterostructure of a few layers of chromium triiodide and single-layer WSe$_2$\cite{surfacelayercri3}. However, the surface spin-polarized charge transfer in bulk Cr-based AFM material was not observed before.

% third figure
 \begin{figure}[!ht]

    \includegraphics[width=0.98\textwidth]{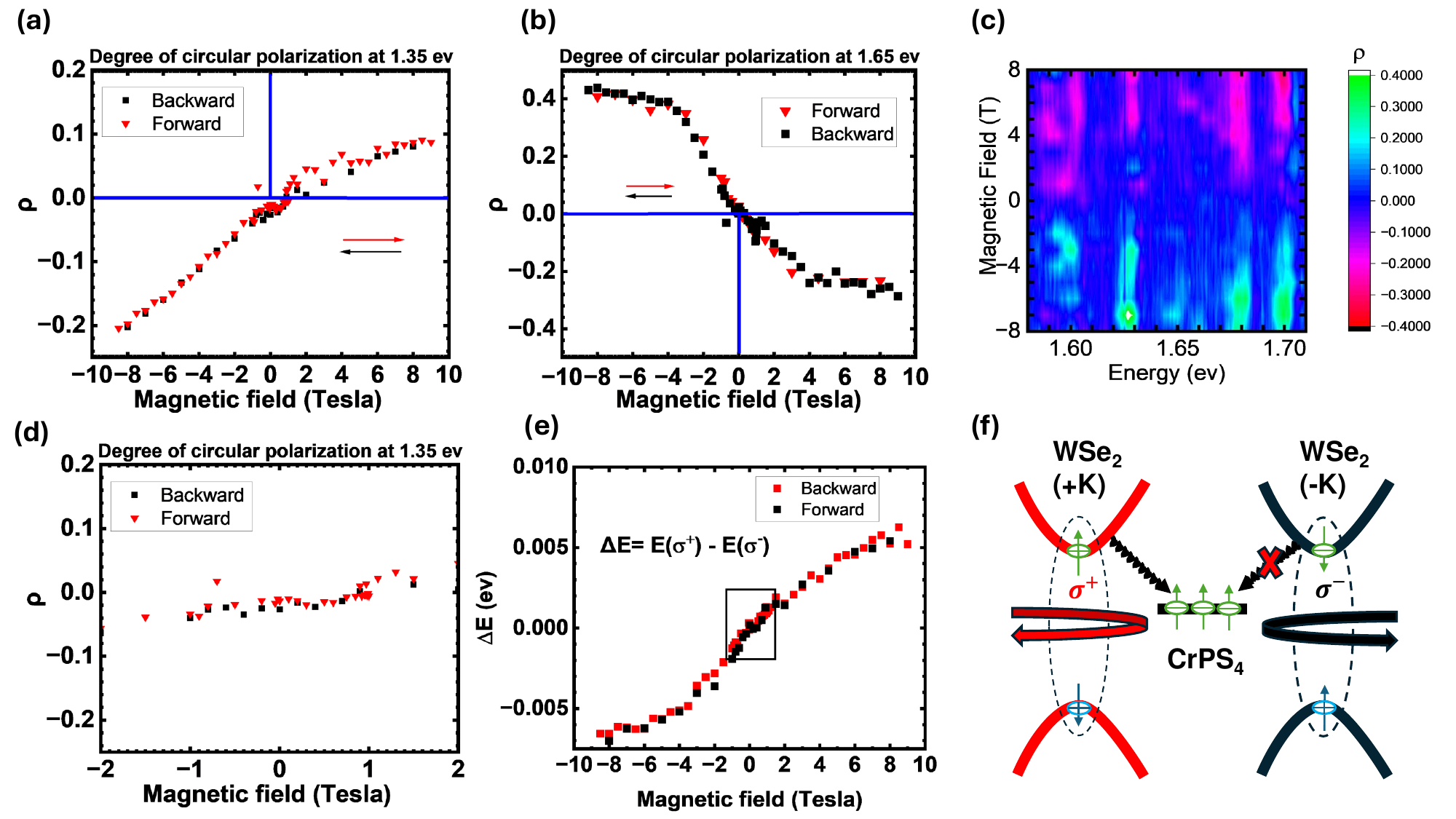}
    \caption{The degree of circular polarization ($\rho$) and Zeeman splitting measurements as a function of magnetic fields are shown. (a) illustrates the $\rho$ measurement of one of the peaks of CrPS$_4$ centered around 1.355 eV at varying magnetic fields, with arrows indicating the direction of the magnetic sweep. (b) presents the $\rho$ measurement of the intrinsic defect peak of WSe$_2$ centered around 1.65 eV, while (c) displays a color plot illustrating the $\rho$ of multiple defect-based bound excitons in WSe$_2$. (d) is a zoomed-in plot of (a), focusing on specific details. (e) shows the Zeeman splitting of the defect peak of WSe$_2$ centered around 1.65 eV. Additionally, (f) provides a schematic demonstrating the spin-polarized charge transfer effect in the heterostructure due to type-II band alignment.  } 
\end{figure}

Fig. 3(e) illustrates the Zeeman splitting of the defect peak centered at 1.65 eV. The marked region highlights the linear behavior of Zeeman splitting when CrPS$_4$ is in the A-type AFM state. However, beyond this point, the energy-splitting behavior becomes non-linear and saturates at high magnetic fields. This type of behavior bears resemblance to observations of valley splitting in WSe$_2$ due to MPI with FM substrate europium sulfide,\cite{zeemanWse2} which further confirms the surface layer sensing in bulk CrPS$_4$ and single layer WSe$_2$ heterostructure.

% Fourth figure
\begin{figure}[!ht]

    \includegraphics[width=0.98\textwidth]{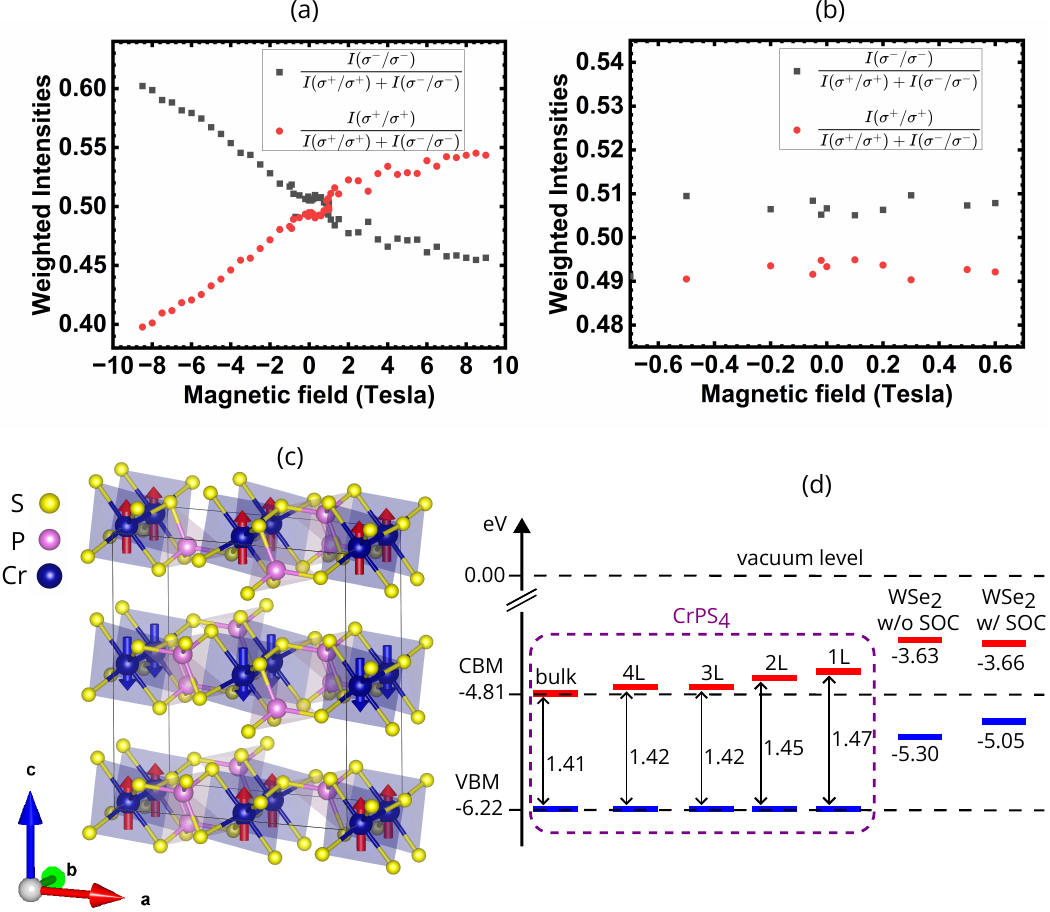}
    \caption{(a) Weighted intensities of the co-polarization component of one of the CrPS$_4$ peaks at 1.355 eV during the backward sweep direction, providing insights into the polarization behavior of the emitted light. (b) A zoomed-in image of (a), focusing specifically on a range corresponding to the A-type AFM state in CrPS$_4$, allowing for a detailed examination of the polarization characteristics within this magnetic regime. (c) The crystal structure of CrPS$_4$, as determined by DFT, offers a visual representation of its atomic arrangement and symmetry properties. (d) Electronic structures of CrPS$_4$ across varying layer thicknesses, ranging from monolayer to bulk, along with comparisons to that of WSe$_2$ with and without spin-orbit coupling, further confirming the type-II band alignment of CrPS$_4$ and WSe$_2$ heterostructure. } 
\end{figure}

After transitioning from the A-type AFM state, CrPS$_4$ enters the C-AFM state, where the magnetic moments align with the b-axis of the crystal\cite{spinflipandafm}. We observe the unidirectional anisotropy in the magnetic sweep measurement in the C-AFM regime. To further understand this behavior we analyze each polarized component of CrPS$_4$ as a function of magnetic fields.

Fig. 4(a) illustrates the weighted intensity of the $\sigma^+$ and $\sigma^-$ polarized transitions of CrPS$_4$ which are used to plot $\rho$ in a backward sweep direction in Fig. 3(a). In the A-type AFM regime, Fig. 3(a) demonstrates a non-zero $\rho$, as depicted in the weighted plot of the polarized component of CrPS$_4$ in Fig. 4(a). Fig. 4(b) provides a zoomed-in view of Fig. 4(a), revealing an unequal yet nearly constant polarized transition from CrPS$_4$ in the A-type AFM regime and exhibiting anisotropic changes in the C-AFM state within positive and negative magnetic field regimes. Although Parity-Time (PT) symmetry suggests uniform polarized component behavior within the A-type AFM regime and isotropic behavior in the C-AFM state, the observed anisotropy may arise from the formation of multiple AFM domains in bulk CrPS$_4$ during the transition from a paramagnetic state to an A-type AFM state upon cooling below its N\'eel temperature. Additionally, introducing WSe$_2$ atop a
magnetic material is known to introduce a strong Rashba effect that significantly alters the magnetic surface textures\cite{dolui2020spin} Further experimental studies are necessary to comprehend this behavior, such as investigating the behavior of $\rho$ after field-cooled magnetization and varying CrPS$_4$ thickness or without WSe$_2$ atop. This unequal but nearly constant polarized transition holds promise for future ultrafast AFM-based memory devices.

The polarized component of CrPS$_4$ PL exhibits sensitivity to laser power. We conducted $\rho$ measurements at various excitation power levels of linearly polarized light at $-8$ Tesla, shown in S3. We observed a nonlinear decrease in $\rho$ with increasing laser power. This behavior can be attributed to spin fluctuations in CrPS$_4$ induced by the rise in temperature due to the increase in laser power, a phenomenon previously observed in this material using NV magnetometry measurements \cite{spinfluctuation}.

We note that the behavior of $\rho$ in both CrPS$_4$ and intrinsic defects in WSe$_2$ remains independent of excitation polarization, as demonstrated in S4. Plots in S4(a) and S4(b) illustrate $\rho$ measurements under both linear and co-polarization excitation of CrPS$_4$ and WSe$_2$ intrinsic defects states, revealing that $\rho$ remains unaffected by excitation polarization. This behavior is attributed to the anisotropic monoclinic crystal structure inherent to these materials \cite{anisotropycrystal}.

In addition to experimental analysis, we also performed DFT calculations to gain better insights into the electronic structure of both CrPS$_4$ and its heterostructure with WSe$_2$. As shown in Fig. 4(c), CrPS$_4$ exhibits a monoclinic crystal structure with anisotropy along the crystal axes directions a and b, resulting in ferromagnetic ground states within each layer, and antiferromagnetically coupled interactions between adjacent layers. The magnetic moment per Cr atom in a single layer is found to be 2.85 $\mu_{\rm B}$, consistent with the previously calculated value\cite{magneticgroundstate}. The b vector is normal to the ac plane while the a and c vectors are at an angle of approximately 92 degrees, owing to the monoclinic crystal structure with C$_2$ rotational symmetry of CrPS$_4$. 

S5 (a) shows the calculated spin-resolved crystal structure where each CrPS$_4$ layer exhibits a net out-of-plane magnetic moment antiferromagnetically coupled with the neighboring layer resulting in CrPS$_4$ being of the A-type AFM structure with a N\'eel temperature of approximately 36 K consistent with our experimental and previously reported results\cite{atypeafmairstable,magneticgroundstate,spinflipandafm}. 

Further, we present the band diagram of bulk CrPS$_4$ with and without spin-orbit coupling, showing no significant changes as shown in S5 (c) and S5 (d).  The negligible spin-orbit coupling in CrPS$_4$ explains why the Zeeman splitting in CrPS$_4$ optical transitions in our experimental data is not resolvable, making the optical transitions in CrPS$_4$ localized. 

Fig. 4d displays the calculated band gap of CrPS$_4$ from a few layers to bulk, closely matching the experimentally detected values. Additionally, we calculated the band gap of WSe$_2$ with and without spin-orbit coupling. In both cases, the band alignment between CrPS$_4$ and WSe$_2$ is type-II, further supporting our experimental findings. 

Furthermore, we calculated the band diagram of a single layer of CrPS$_4$ and WSe$_2$(S5 (f) and S5 (g)), where the highlighted WSe$_2$ and CrPS$_4$ bands illustrate the type-II band alignment between the CrPS$_4$ and WSe$_2$ heterostructure. Detailed information regarding the DFT calculations can be found in the supplementary information.

\section{Conclusion}
In conclusion, we present a method for analyzing the A-type AFM and C-AFM states in optically active 2D bulk semiconductor magnets. Additionally, we successfully detected the surface layer magnetic moment of CrPS$_4$ by utilizing WSe$_2$ defect states in a magnetic heterostructure. Our prediction of type-II band alignment through DFT calculations supports the observed spin-polarized charge transfer in the heterostructure of bulk CrPS$_4$ and single-layer WSe$_2$. Due to its anisotropic monoclinic structure, the spin-polarized charge transfer in heterostructures remains independent of excitation polarization. Overall, our work highlights the potential of utilizing optically active bulk 2D AFM material as ultrafast optical memory storage devices by using its surface layer properties. Furthermore, owing to spin-polarized charge transfer, TMDC-based heterostructures emerge as promising candidates for quantum information processing.

\section{Method}
\subsection{Heterostructure preparation}
For preparing the heterostructure, CrPS$_4$, WSe$_2$, and hexagonal boron nitride (h-BN) flakes were cleaved out from the bulk using less residual tape. Then the flakes are further cleaved into thinner portions using the tape multiple times. For CrPS$_4$, the thinner flakes in the tape are pressed into the Si/SiO$_2$ substrate and for WSe$_2$ and h-BN the flakes are pressed in Polydimethylsiloxane (PDMS) after removing the tape multiple flakes are transferred into the Si/SiO$_2$ substrate and PDMS. The bulk CrPS$_4$ and single layer WSe$_2$ were identified by optical microscope using optical contrast from the flake and further confirmed by Raman and PL spectroscopy. The heterostructure was prepared by stacking both flakes together with h-BN encapsulation on top and bottom using PC and transferred into the marked substrate for PL measurements.   

\subsection{Magneto-photoluminescence}
The sample was positioned inside the Attocube-2100 cryostat with the superconducting magnet in Faraday geometry range up to $\pm$ 9 Tesla, equipped with a confocal microscope head placed on top of the cryostat to conduct polarized resolved photoluminescence measurements at 1.8 K and 300 K temperature. The cryostat objective, featuring a numerical aperture (NA) of 0.82, was utilized to facilitate the excitation of the sample using a 532 nm continuous wave laser. The cryostat objective provided a spot size of 791.5 nm, ensuring accurate focusing of the laser beam onto the sample. Circularly polarized light ( $\sigma$$^+$ and $\sigma$$^-$) was generated and detected by the combination of linear polarizer and quarter wave-plate from thorlabs.  To capture the emitted light, the collection arm of the microscope was connected to a fiber-optic cable, which directed the light to the entrance port of the Teledyne HRS-750 triple-gating imaging spectrometer. The spectrometer, with a focal length of 750 mm, enabled efficient spectral analysis. Subsequently, the collimated light was diffracted from the grating within the spectrometer and directed toward a nitrogen-cooled CCD camera. The camera featured a pixel array configuration of 1340x400, which facilitated precise and detailed spectral measurements.
\begin{suppinfo}

Computational method for DFT calculation, room temperature co-polarization resolved PL of CrPS$_4$ and WSe$_2$ heterostructure, degree of circular polarization measurement of CrPS$_4$ and WSe$_2$ at $\pm 8.5$ Tesla, power-dependent degree of circular polarization measurement at $- 8$ Tesla, magnetic sweep measurement of heterostructure with excitation of different polarization and electronic structure of bulk CrPS$_4$ and heterostructure between its monolayer and WSe$_2$.

\end{suppinfo}

\subsection{Contribution}
The measurements and characterization were done by M.H.S. The device fabrication was done by M.H.S and M.W. The theoretical calculations were performed by D.Q.H, supervised by A.J. J.X, YR, CC and M.H.S contributed to the interpretation of data. C.C supervised the project and wrote the draft together with M.H.S. All authors contributed to the manuscript preparation and approved its final version.

% \begin{acknowledgement}

% Please use ``The authors thank \ldots'' rather than ``The
% authors would like to thank \ldots''.

% The author thanks Mats Dahlgren for version one of \textsf{achemso},
% and Donald Arseneau for the code taken from \textsf{cite} to move
% citations after punctuation. Many users have provided feedback on the
% class, which is reflected in all of the different demonstrations
% shown in this document.

% \end{acknowledgement}
\subsection{Acknowledgement}
The work was primarily supported as part of the Center for Hybrid, Active, and Responsive Materials (CHARM) funded by the National Science Foundation (NSF) DMR-2011824 Seed Award program. C.C. acknowledges partial support by the  NSF Award OIA-2217786. The authors thank Xi Wang from the Department of Material Science and Engineering, University of Delaware for letting us use their transfer station and room temperature PL and Raman measurement setup. The authors thank Yi Ji from the Department of Physics and Astronomy, University of Delaware for letting us use their optical microscope.

\bibliography{acs-achemso.bib}

\section{Supplementary Information}
\subsection{Computational Method}
The first-principles calculations were conducted using density functional theory (DFT) with periodic boundary conditions, employing the projector augmented wave (PAW) method as implemented in the Vienna Ab initio Simulation Package (VASP). The 3p, 3d, and 4s orbitals of Cr, the 3s and 3p orbitals of P, the 3s and 3p orbitals of S, the 4s, 4p, and 4d orbitals of W, and the 3s and 3p orbitals of Se were explicitly considered as valence shells in the PAW potentials. The Kohn-Sham wave functions were expanded using plane-wave basis sets with a kinetic energy cutoff of 500 eV. For optimizing geometrical structures and self-consistently determining electronic properties, we employed the strongly constrained and appropriately normed (SCAN) meta-generalized gradient approximation for short- and intermediate-range interactions, combined with the long-range van der Waals (vdW) interactions from the revised Vydrov–van Voorhis (rVV10) non-local correlation functional, collectively referred to as SCAN-rVV10. This choice of exchange-correlation function was shown to perform well for layered materials.

Single layers of CrPS$_4$ and WSe$_2$ were calculated without vdW correction, while bulk and few-layer structures of CrPS$_4$, as well as heterostructures consisting of sandwiched single layers of CrPS$_4$ and WSe$_2$, were simulated using a slab approach including vdW. A vacuum space of at least 20 Å along the normal plane of the slab was employed to eliminate spurious interactions between slab images. For structural optimization, $\Gamma$-centered $k$-point meshes of 3×5×3 for bulk CrPS$_4$ and 3×5×1 for few-layer structures were used to sample the Brillouin zones; forces on each atom were self-consistently converged to be smaller than 0.005 eV/\AA. For electronic structure calculations, denser $k$-point grids of 5×7×4 and 5×7×1 were used for bulk and few-layer structures, respectively. The optimized lattice constants of bulk (with a 40-atoms unit cell) are a = 10.858 \AA, b = 7.259 \AA, and c = 12.292 \AA, where vector b is perpendicular to both a and c, while a and c are at an angle of approximately 92$^0$. The calculated local magnetic moment 2.85 $\mu_{\rm B}$ per Cr atom, is in good agreement with experimental measurements.

To determine the band alignment between CrPS$_4$ and WSe$_2$ in heterostructures, electronic levels of all materials were aligned with respect to a common reference, which is the vacuum level.

 \vspace{100pt}

% Reset figure counter and redefine figure numbering
\setcounter{figure}{0}
\renewcommand{\thefigure}{S\arabic{figure}}

\section{Supplementary figure 1}
% Your Supplementary_figures 1
\begin{figure}[!hb]
    \includegraphics[width=0.95\textwidth]{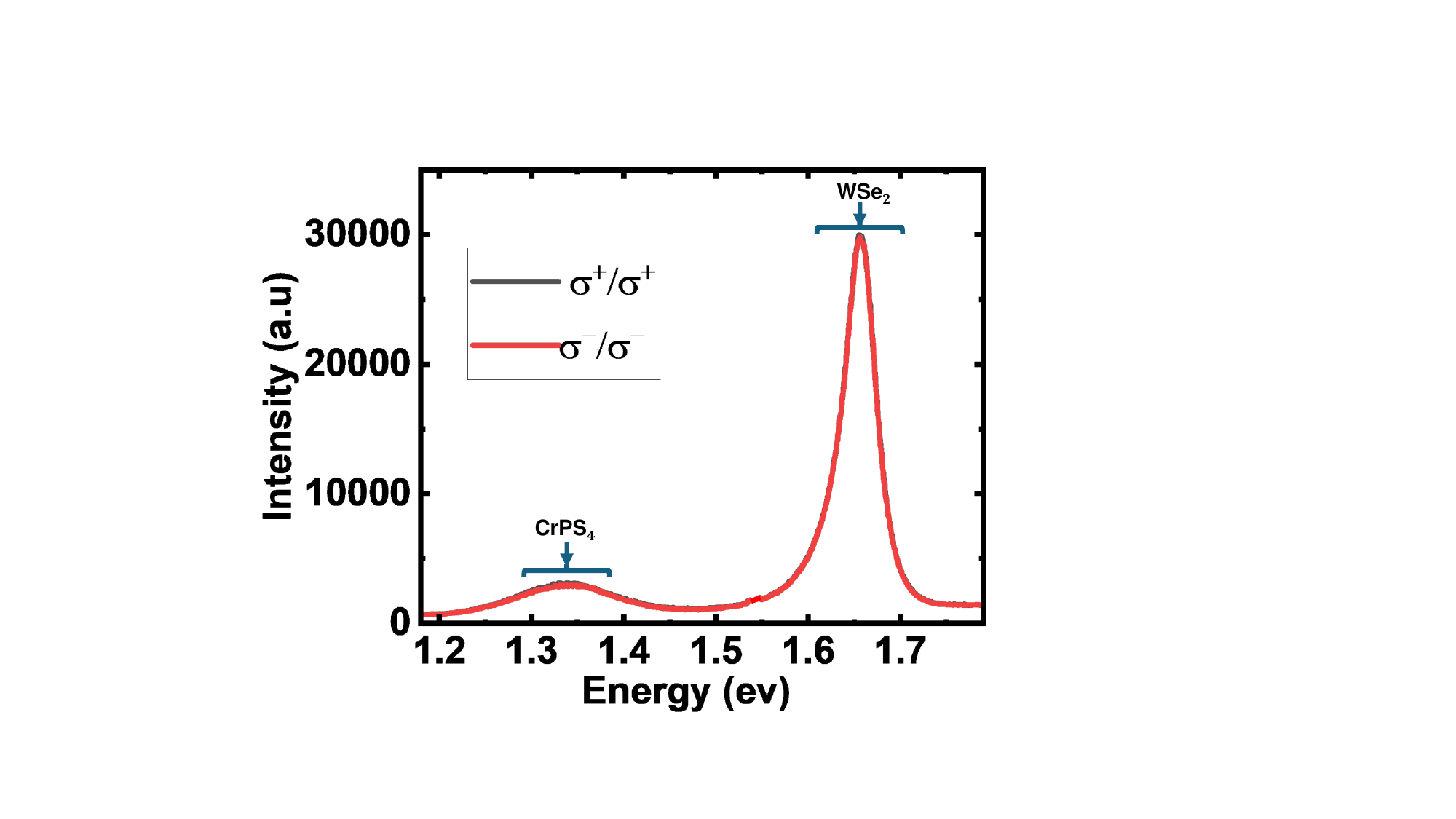}
    \caption{ The co-polarization resolved photoluminescence (PL) of the heterostructure at room temperature is shown, the peak centered around ~1.35 eV corresponds to bulk CrPS$_4$, while the peak centered around ~1.65 eV is attributed to the excitonic transition in WSe$_2$. Notably, the negligible polarization contrast and energy degeneracy observed in the excitonic transition indicate the absence of a magnetic proximity effect. This lack of effect can be attributed to the paramagnetic behavior exhibited by CrPS$_4$ at room temperature.    }
\end{figure}
 \vspace{200pt}

\section{Supplementary figure 2}
% Your Supplementary_figures 2
\begin{figure}[!hb]
    \includegraphics[width=1\textwidth]{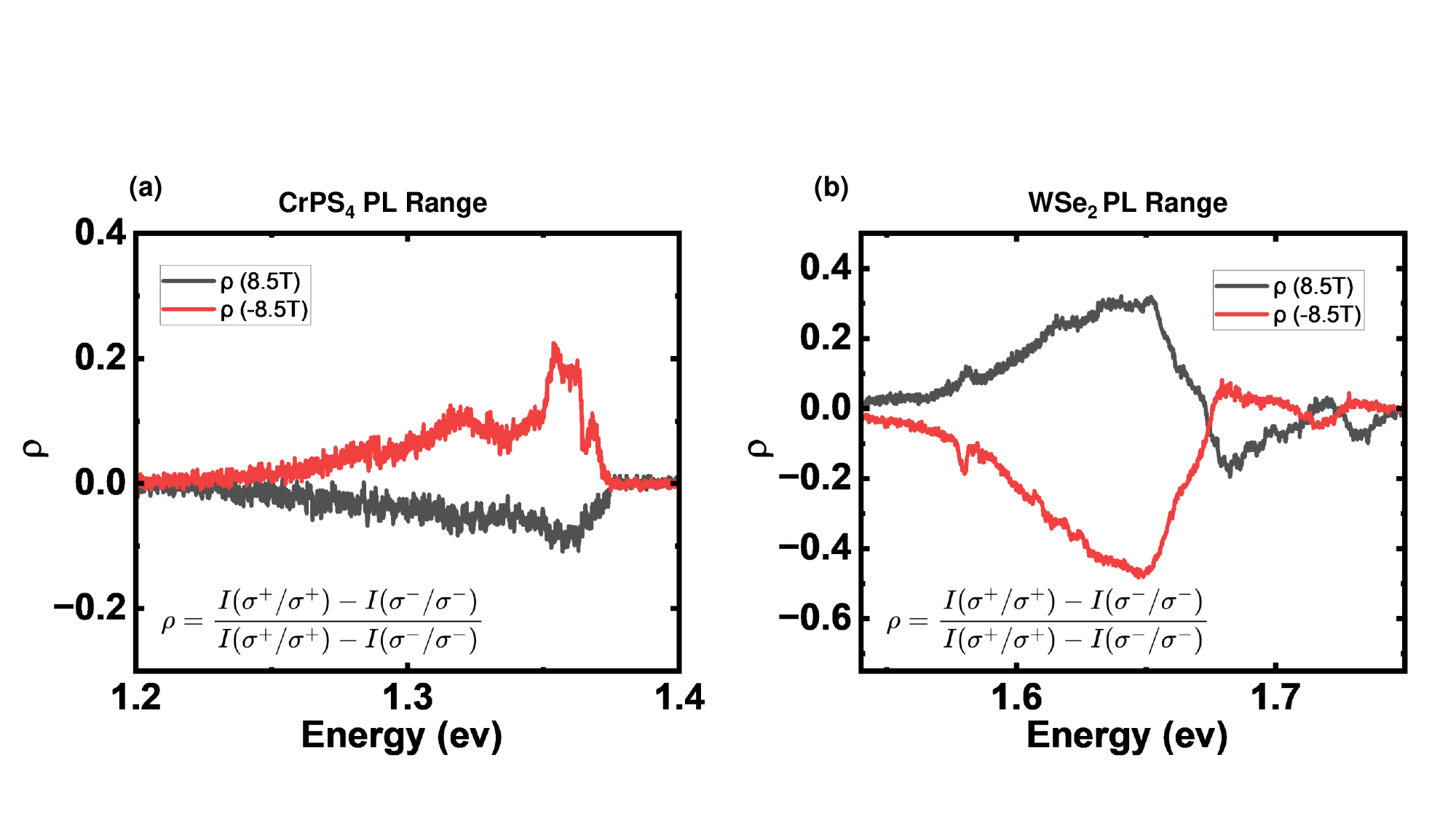}
    \caption{Degree of circular polarization measurement at $\pm$ 8.5 Tesla: (a) illustrates the degree of circular polarization in the PL range of CrPS$_4$. All peaks of CrPS$_4$ exhibit a positive (negative) degree of circular polarization behavior at $+$8.5 ($-$8.5) Tesla. (b) depicts the degree of circular polarization measurement in the PL range of WSe$_2$, where the peaks centered around 1.65 eV and 1.725 eV corresponding to defect and excitonic transitions show the opposite behavior compared to CrPS$_4$ at $+$8.5 ($-$8.5) Tesla. However, certain regions (peaks at 1.68 eV and 1.73 eV) display similar behavior to CrPS$_4$ due to the Zeeman splitting of excitonic transitions, as depicted in Figures 2(d) and 2(e).      }
\end{figure}
\vspace{200pt}
\section{Supplementary figure 3}
% Your Supplementary_figures 3
\begin{figure}
    \includegraphics[width=0.95\textwidth]{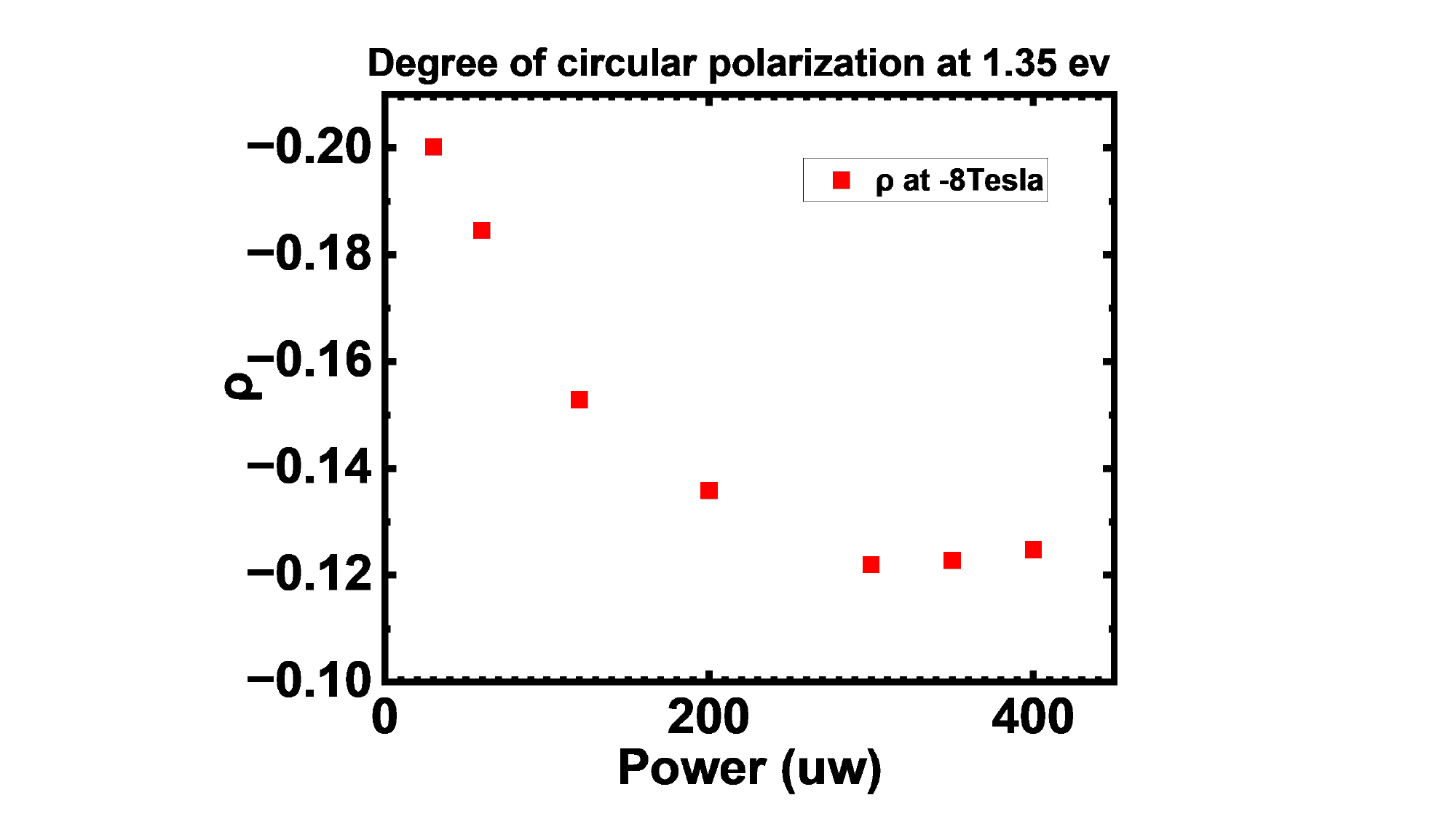}
    \caption{Power-dependent degree of circular polarization of CrPS$_4$ at -8 Tesla: We conduct the degree of circular polarization measurement at various laser excitation powers. Our observations reveal a non-linear behavior in the degree of circular polarization, which decreases non-linearly with an increase in the excitation power. }
\end{figure}

\vspace{250pt}
\section{Supplementary figure 4}
% Your Supplementary_figures 4
\begin{figure}
    \includegraphics[width=0.95\textwidth]{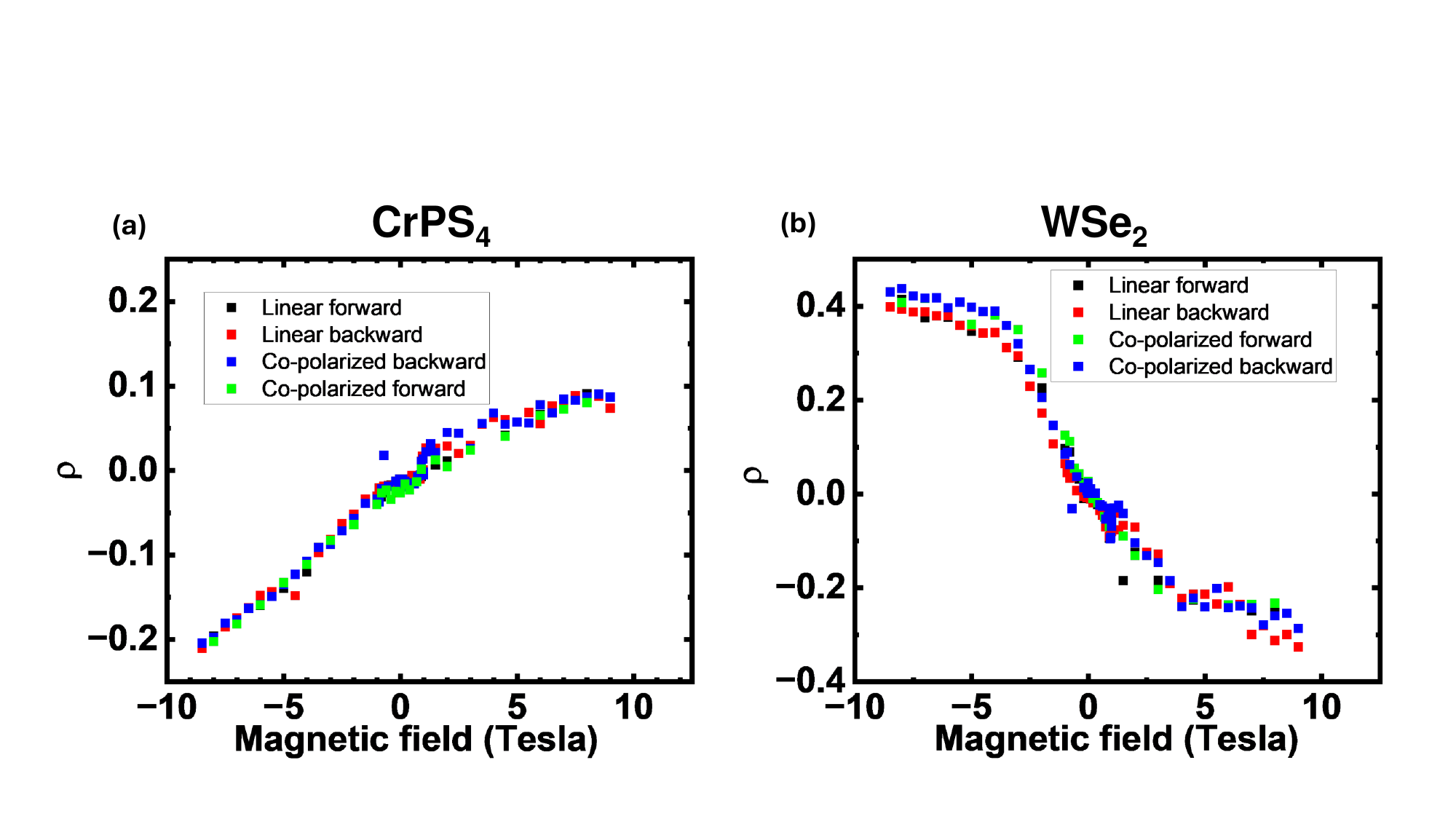}
    \caption{Magnetic sweep measurement of heterostructure with excitation of different polarizations: We conduct a degree of circular polarization measurement at various magnetic fields using light with different polarizations. Our findings indicate that the behavior of spin-polarized charge transfer between CrPS$_4$ and WSe$_2$ remains consistent regardless of the excitation light's polarization.}
\end{figure}

\vspace{205pt}
\section{Supplementary figure 5}
% Your Supplementary_figures 5
\begin{figure}
    \includegraphics[width=0.92\textwidth]{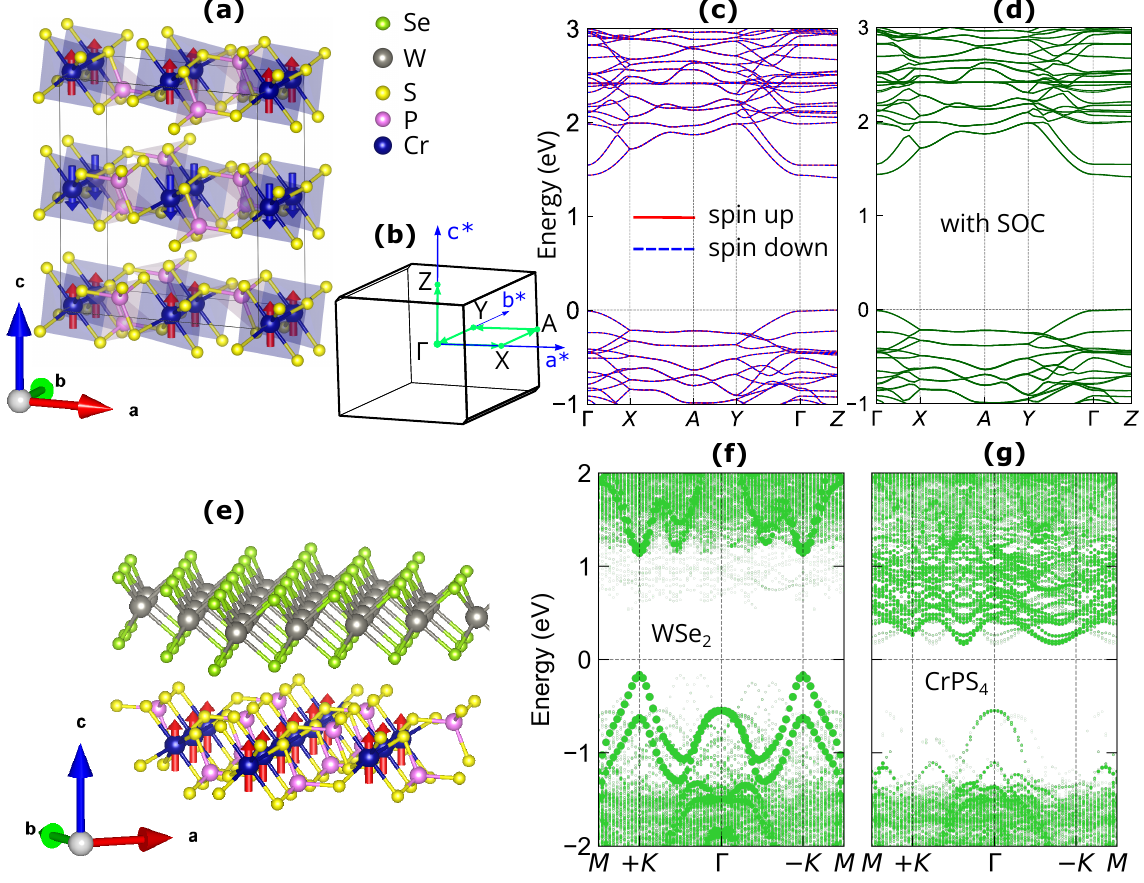}
    \caption{Electronic structures of bulk CrPS$_4$ and heterostructure between its monolayer and WSe$_2$. (a) The A-type antiferromagnetic AFM crystal structure of CrPS$_4$, where each layer possesses a net out-of-plane magnetic moment that is antiferromagnetically coupled with the neighboring layer; each Cr atom within a single layer exhibits a magnetic moment of 2.85 $\mu_{\rm B}$. (b) The Brillouin zone of bulk CrPS$_4$ shown in (a). (c) and (d) Electronic band structures of CrPS$_4$ without (spin-polarized) and with the inclusion of spin-orbit coupling (SOC), respectively; no significant change was found in the band structure when including SOC. (e) The heterostructure from monolayer of CrPS$_4$ and WSe$_2$. Unfolding band structure of the heterostructure and its projection on the material's components, depicting the type-II band alignment between a single layer of CrPS$_4$ and a single layer of WSe$_2$; the band diagrams highlight the contribution of (f) WSe$_2$ and (g) CrPS$_4$.}
\end{figure}

\end{document}